\documentclass[conference]{IEEEtran}
\IEEEoverridecommandlockouts
\usepackage{cite}
\usepackage{amsmath,amssymb,amsfonts}
\usepackage{algorithmic}
\usepackage{algorithm}
\usepackage{graphicx}
\usepackage{textcomp}
\usepackage{xcolor}
\def\BibTeX{{\rm B\kern-.05em{\sc i\kern-.025em b}\kern-.08em
    T\kern-.1667em\lower.7ex\hbox{E}\kern-.125emX}}
\begin{document}
	
	\title{Digital Self-Interference Cancellation With Robust Multi-layered Total Least Mean Squares Adaptive Filters\\
	}
	
\author{\IEEEauthorblockN{Shiyu Song\textsuperscript{1}, Yanqun Tang\textsuperscript{1,*}, Xizhang Wei\textsuperscript{1}, Yu Zhou\textsuperscript{1}, Xianjie Lu\textsuperscript{1}, Zhengpeng Wang\textsuperscript{1}, Songhu Ge\textsuperscript{2}}
\IEEEauthorblockA{
\textit{\textsuperscript{1}School of Electronics and Communication Engineering, Sun Yat-sen University, Shenzhen, China} \\  
 \textit{\textsuperscript{2}National Key Laboratory of Electromagnetic Energy, Naval University of Engineering, Wuhan, China }\\ Corresponding Author: tangyq8@mail.sysu.edu.cn*} }
	\maketitle
	
	\begin{abstract}
In simultaneous transmit and receive (STAR) wireless communications, digital self-interference (SI) cancellation is required before estimating the remote transmission (RT) channel. 
Considering the inherent connection between SI channel reconstruction and RT channel estimation, we propose a multi-layered M-estimate total least mean squares (m-MTLS)  joint estimator to estimate both channels. 
In each layer, our proposed m-MTLS estimator first employs an M-estimate total least mean squares (MTLS) algorithm to eliminate 
residual SI from the received signal and give a new estimation of the RT channel. 
Then, it gives the final RT channel estimation based on the weighted sum of the estimation values obtained from each layer.  
Compared to traditional minimum mean square error (MMSE) estimator and single-layered MTLS estimator, it demonstrates that the m-MTLS estimator has better performance of normalized mean squared difference (NMSD). 
Besides, the simulation results also show the robustness of m-MTLS estimator even in scenarios where the local reference signal is contaminated with noise, and the received signal is impacted by strong impulse noise. 
	\end{abstract}


	\section{Introduction}
Simultaneous transmit and receive (STAR) system enables the transceiver to transmit and receive signals simultaneously at the same carrier frequency. Because of its theoretical benefits of doubled spectral efficiency compared to traditional half-duplex systems, it is considered a promising approach to increase data throughput. However, it will generate a self-interference (SI)  that is 100-120dB stronger than the desired signal due to direct leakage or coupling between the transmit and receive antennas which are physically close to each other\cite{b0}. Therefore, eliminating strong SI is currently a significant challenge in STAR communication systems.

To improve the self-interference cancellation (SIC) capability, various algorithms based on adaptive filter in the digital domain have been developed. The authors in \cite{b1} perform a fitting of link characteristics using a linear finite impulse response filter and conduct a simulation analysis of interference cancellation by employing the least mean squares (LMS) adaptive filter. The results show a cancellation ratio exceeding 20dB. A fractional order LMS  method is proposed to achieve a better SIC performance in linear frequency-modulated continuous wave radar in \cite{b2}. A comparison of various adaptive filtering strategies for SIC in Long-Term Evolution communication systems is presented in \cite{b3}. 
In the STAR communication system, our objective is to estimate the remote transmission (RT) channel after eliminating SI for the purpose of equalizing SI-suppressed signals to detect data symbols from the signal of interest.
Therefore, the authors use the joint channel estimation to eliminate SI  and estimate the RT channel with recursive least squares (RLS) at the same time in \cite{b6}. It is worth noting that as the interference-to-signal ratio (ISR) increases, the performance may deteriorate due to the extended length of the filter.
To improve the performance of joint channel estimation, particularly in scenarios where the differential power is high, the authors in \cite{b7} employ the multi-layered recursive least squares (m-RLS) estimator. 

However, the methods mentioned above can only achieve an optimal solution when the input data matrix is perfectly accurate.
In the presence of noise in both the data matrix and the observed data, the total least mean squares (TLS) algorithm proposed in \cite{b4} exhibits smaller fitting errors. 
In practical communication environments, there are often additional sources of noise, such as electromagnetic interference, lightning noise, radar signals, and other natural and man-made disturbances. These types of noise typically exhibit strong pulse-like characteristics within very short time durations, which can lead to performance degradation of various filtering algorithms based on the assumption of Gaussian noise 
\cite{7862211}.
To further improve the performance of TLS in the presence of impulse noise, a novel robust adaptive algorithm combining the advantages of TLS and M-estimate function is proposed in \cite{b5} along with its variable step-size version.

In this paper, we combine the advantages of M-estimate total least mean squares 
 (MTLS) algorithm and the multi-layered joint estimator by employing a robust multi-layered M-estimate total least mean squares (m-MTLS) estimator to enhance the performance of the joint channel estimation in scenarios where both the data matrix and the observed data are contaminated by Gaussian noise, while the observation signal is also affected by impulse noise.
This method overcomes the problem of deteriorated cancellation performance in the digital domain SIC of traditional STAR systems at high ISR and exhibits stronger robustness in the presence of impulse noise.
The final simulation results demonstrate that by iteratively performing SIC, the m-MTLS  estimator effectively reduces the convergence error as the ISR is increased from 20dB to 40dB, thereby reducing the convergence error under high ISR conditions. Furthermore, in the presence of impulse noise, the m-MTLS estimator exhibits superior robustness compared to MMSE estimator.


\section{System Model}
\begin{figure}[htbp]
		\centerline{\includegraphics[width=1\linewidth]{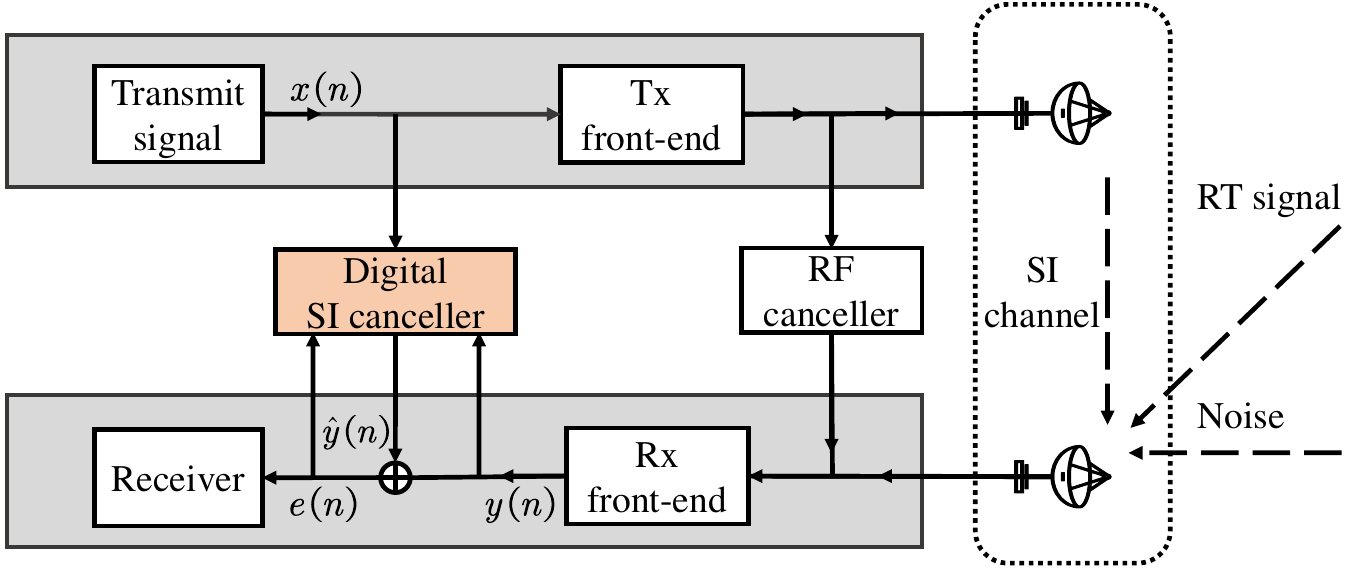}}
		\caption{ An illustration of STAR system.}
		\label{channel}	
	\end{figure}

	Fig. \ref{channel} shows the framework of STAR system. The received signal after LNA amplification and analog domain SIC is down-converted to baseband signal. At time instant $n$, the received signal $y(n)$ can be represented as
	\begin{equation}
		\begin{aligned}
			y(n)=r(n)+s(n)+v(n), 
		\end{aligned}
		\label{11}
	\end{equation}	
	where $r(n)$ and $v(n)$ denote the desired signal from the remote source and noise  at time instant $n$, respectively.  	The residual SI $s(n)$,  can be represented by a linear model as
	
		\begin{equation}
		\begin{aligned}
			s(n)=\mathbf{w}^T(n)\mathbf{i}(n),
		\end{aligned}
		\label{12}
	\end{equation}		
	where $\mathbf{w}\in \mathbb{R} ^{N\times 1}$ denotes the SI channel and $\mathbf{i}(n)$ represents the known local signal.

Based on the aforementioned analysis, in order to eliminate the residual SI in the digital domain, we only need to obtain the optimal SI channel response estimate $\mathbf{\hat{w}}(n) $ for $\mathbf{w}(n)$ and subtract the reconstructed SI from $y(n)$. After SIC, the residual signal can be considered as the RT signal, and further processing, such as RT channel estimation and equalization can be performed to detect data symbols from the RT signal.




\section{Proposed Joint \lowercase{m}-MTLS Estimator}
In traditional STAR systems, it is common first  to eliminate SI  from the received signal and then estimate and equalize the filtered signal.
However, when the ISR is large, even small errors can result in significant residual SI. 	
Moreover, LMS and other basic adaptive algorithms can not achieve an optimal solution when the local reference signal is contaminated by noise.
	\subsection{Total Least Squares Problem}
	 We consider a linear model where both the input and output signals are contaminated by noise, which can be illustrated by the following equation
	\begin{equation}
		(y_n+v_n)=(\mathbf{x}_n+\mathbf{u}_n)^T \mathbf{h},
		\label{1}
	\end{equation}
	where $\mathbf{h}\in\mathbb{R}^{L\times 1}$ is the unknown system vector to be estimated, $\mathbf{x}_n\in \mathbb{R} ^{L\times 1}$ and $y_n\in \mathbb{R} $ are the input vector and output signal at time $n$, respectively. The vector $\boldsymbol{u}_n \sim \,\,N\left( \mathbf{0},\sigma _{i}^{2}\mathbf{I} \right)  \in R^{L \times 1}$ and  $v_n \sim \,\,N\left( 0,\sigma _{o}^{2} \right) \in R$ stand for input noise and output noise with $\mathbf{0}$ and $\mathbf{I}$ denoting the all-zero column vector and identity matrix, respectively. 
	
	The solution to the TLS problem can be obtained by performing the direct eigenvalue decomposition (EVD) of the augmented covariance matrix, but this process  requires a significant computational effort \cite{b4}. Therefore, the authors in \cite{b8} propose an iterative scheme to find the TLS fit by minimizing the following Rayleigh quotient cost function, which can be formulated as
	\begin{equation}
		\min _{\boldsymbol{w}} J(\boldsymbol{w})=\frac{1}{N} \sum_{n=1}^N \frac{\left(\tilde{y}_n-\boldsymbol{w}^T \tilde{\boldsymbol{x}}_n\right)^2}{\|\boldsymbol{w}\|^2+\gamma},
		\label{a}
	\end{equation}
	where $\gamma \overset{\text{def}}{ =}\sigma _{o}^{2}/\sigma _{i}^{2}$ is a  parameter to normalize the noise variances, $\tilde{y}_n = y_n+v_n $ and $\mathbf{\tilde{x}}_n = \mathbf{x}_n+\mathbf{u}_n $. 
	
	We can replace the time average in (\ref{a}) with the expectation operation and 
	the expression can be rewritten as
	\begin{equation}
		\min _{\boldsymbol{w}}J(\boldsymbol{w})=E\left[\frac{\left(\tilde{y}_n-\boldsymbol{w}^T \tilde{\boldsymbol{x}}_n\right)^2}{\|\boldsymbol{w}\|^2+\gamma}\right]=E\left[\frac{e^2_n}{\|{\boldsymbol{w}}\|^2+\gamma}\right].
	\end{equation}
	
	Then we can obtain the optimal solution of the TLS by updating the weights of the filter according to the following equation with gradient-descent method.
	\begin{equation}
		\begin{aligned}
			\mathbf{w}_{n+1} & =\mathbf{w}_n-\mu \hat{\mathbf{g}}\left(\mathbf{w}_n\right) \\
			& =\mathbf{w}_n+ \mu \alpha_n\left(\tilde{\mathbf{x}}_n+\alpha_n \mathbf{w}_n\right),
		\end{aligned}
	\end{equation}
	where $\mu$ is the step-size parameter, and $\hat{\mathbf{g}}$ is  the instantaneous gradient of $J(\boldsymbol{w})$, which can be represented as
	
	\begin{equation}
		\begin{aligned}
			&\hat{\mathbf{g}}(\mathbf{w})=\alpha_n\left(\tilde{\mathbf{x}}_n+\alpha_n \mathbf{w}_n\right),\\
			&\alpha_n \overset{\text{def}}{=}\frac{\tilde{y}_n-\tilde{\mathbf{x}}_n^T \mathbf{w}_n}{\left\|\mathbf{w}_n\right\|^2+\gamma}.
		\end{aligned}
	\end{equation}

	\subsection{M-estimate Function}
	The TLS method considers the optimal solution under Gaussian noise. However, when the signal is affected by impulsive noise, traditional TLS estimation may not yield effective results.  According to \cite{b9}, we can use M-estimate function to improve the robustness of TLS. The new cost function in \cite{b10} can be written as 
	
	\begin{equation}
		\min _{\boldsymbol{w}}J(\boldsymbol{w})=E\left[\frac{\rho(e^2_n)}{\|{\boldsymbol{w}}\|^2+\gamma}\right],
	\end{equation}
	where $\rho(\cdot)$ is the M-estimate function which is a real-valued even function, given by
	\begin{equation}
		\rho(e_n)= \begin{cases}e^2_n / 2, & |e_n|<\xi \\ \xi^2 / 2, & |e_n| \geq \xi\end{cases},
	\end{equation}
	where $\xi=c_1 \hat{\sigma_e}$ is a parameter controlling the threshold for misjudgment, which is taken as $c_1=2.576$ in this case\cite{b11}, and $\hat{\sigma_e}$ can be calculated by
	\begin{equation}
		\hat{\sigma}_{e_n}^{2}=\lambda _{\sigma}\hat{\sigma}_{e_{n-1}}^{2}+c_2\left( 1-\lambda _{\sigma} \right) \text{med}\left( A_{e_n} \right), 
	\end{equation}
	where $\boldsymbol{A}_e(n)=\left[e^2(n), e^2(n-1), \ldots, e^2\left(n-N_w+1\right)\right]$, $\text{med} (\cdot)$ is the median operator\cite{b12}, $\lambda_\sigma$ is usually chosen between 0.98 and 0.99,  $c_2=1.483\left(1+5 /\left(N_w-1\right)\right)$, and $N_w$ is the length of the estimation window.
	
	Taking the partial derivative of the new cost function, we can obtain
	\begin{equation}
		\hat{\boldsymbol{g}}(\boldsymbol{w})= \begin{cases}-\frac{ \left(\|{\boldsymbol{w}}\|^2+\gamma\right) e_n \tilde{\boldsymbol{x}}_n+e^2_n \boldsymbol{w}}{ \left(  \| \boldsymbol{w}  \|^2 +\gamma \right)^2}, & |e_n|<\xi \\ 0, & |e_n| \geq \xi\end{cases},
	\end{equation}
	and
	\begin{equation}
		\begin{aligned}
			\mathbf{w}_{n+1} & =\mathbf{w}_n-\mu \hat{\boldsymbol{g}}\left(\mathbf{w}_n\right). 
		\end{aligned}
	\end{equation}
	
	\begin{figure}[htbp]
		\centerline{\includegraphics[width=0.8\linewidth]{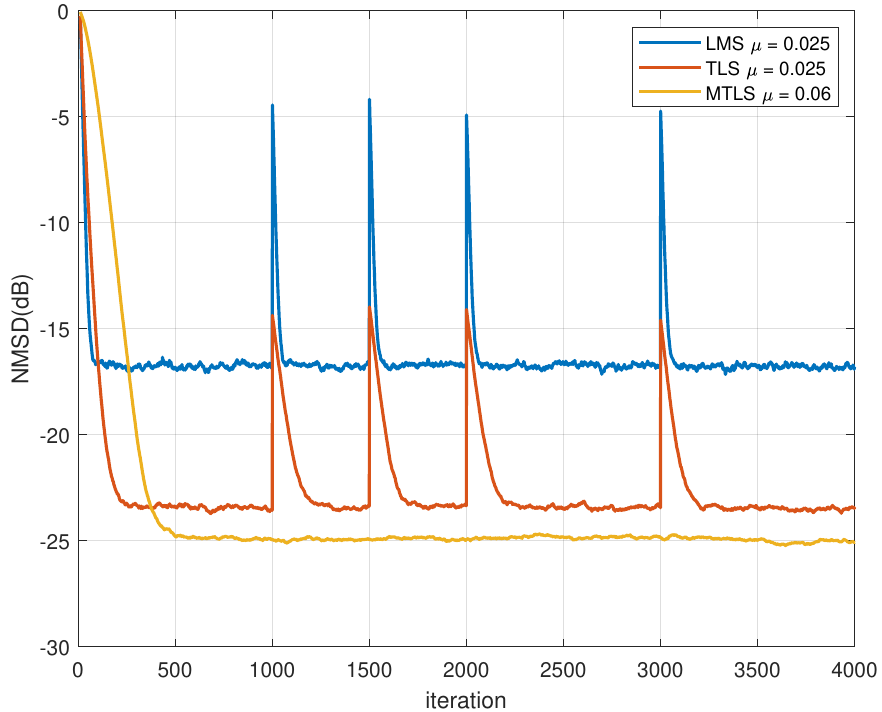}}
		\caption{Average NMSD learning curve with input noise $v_in = v_b$ and output noise $v_{out}= v_a+v_i$, where $\sigma_a^2 = \sigma_b^2=0.1$ and $\sigma_i^2=10$. }
		\label{fig1}
	\end{figure}
	
	Assuming the unknown vector $\mathbf{h}$ in (\ref{1}) has a length of $L=10$ and satisfies $\|\mathbf{h}\|^2=1$, the input noise is given by $v_{in}=v_b$, and the output noise is represented by $v_{out} = v_a+v_i$. Here, $v_a$ and $v_b$ follow a Gaussian distribution with a variance of $\sigma_a^2 = \sigma_b^2=0.1$, while $v_i \sim N(0,10)$ represents impulsive noise occurring at time instants $n = 1000, 1500, 2000$, and $3000$. We apply the LMS, TLS, and MTLS algorithms for filtering, and the performance of these algorithms is evaluated using the normalized mean squared difference (NMSD), defined as NMSD $= 10\log\left(\|\mathbf{w}_n-\mathbf{h}\|^2 / {\|\mathbf{h}\|^2}\right)$. Here, normalization is performed to facilitate a better comparison of algorithm performance across different SNR and ISR settings.
	
	The convergence curves are shown in Fig. \ref{fig1}. The LMS and TLS exhibit significant peaks at  $n = 1000, 1500, 2000$, and $3000$ due to noise interference, while the MTLS remains unaffected by such interference.
	
		\begin{figure*}[htbp]
	\centerline{\includegraphics[width=1\linewidth]{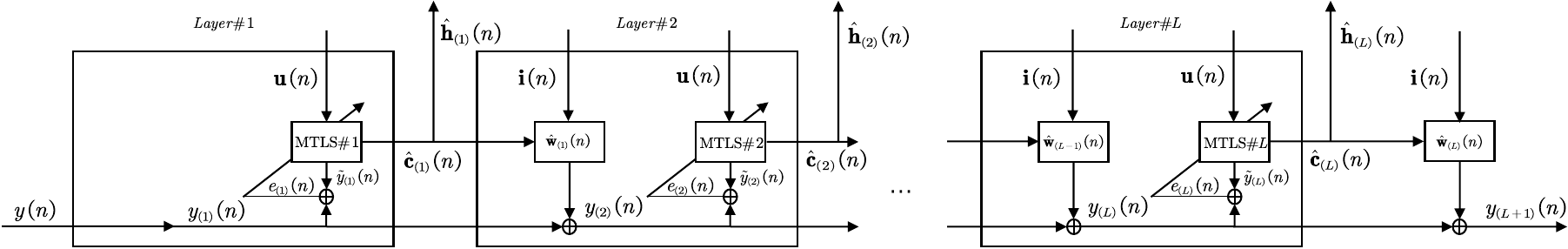}}
	\caption{The structure of m-MTLS joint channel estimator.}
	\label{mmtls}
\end{figure*}	

\subsection{\lowercase{m}-MTLS Estimator}
		
In traditional SIC cases, improving the SNR alone is insufficient to accurately estimate the RT channel.
	 Therefore, we employ a joint estimator to enhance the estimation accuracy in high-SNR scenarios. Additionally, multiple layers of filtering are used to improve performance in high-ISR conditions. 	Thus, (\ref{11}) and (\ref{12}) can be rewritten in \cite{b7} as

\begin{equation}
	\begin{aligned}
		y(n)=\mathbf{w}^T(n)\mathbf{i}(n)+\mathbf{h}^T(n)\mathbf{x}(n)+v(n), 
	\end{aligned}
	\label{12}
\end{equation}
where $r(n)=\mathbf{h}^T(n)\mathbf{x}(n)$, in which $\mathbf{h}\in \mathbb{R} ^{M\times 1}$ denotes the RT channel response, and $\mathbf{x}(n)$ represents the transmitted signal from the remote terminal.

By merging the two signals into a longer vector, according to equation (\ref{12}), we can obtain the following expression
\begin{equation}
	\begin{aligned}
		y_{(1)}(n)=\mathbf{c}^T(n)\mathbf{u}(n)+v(n),
	\end{aligned}
\end{equation}		
where $\mathbf{c}^T(n)=[\mathbf{w}^T(n), \mathbf{h}^T(n)] \in \mathbb{R} ^{1 \times (N+M)}$ combines the coefficients of both the SI and RT channels, which is used for estimating the  new channel, and $\mathbf{u}(n)=[\mathbf{i}^T(n), \mathbf{x}^T(n)]^T$ is the corresponding input vector. 

The framework of the devised m-MTLS filter is depicted in Fig. \ref{mmtls}.
At the first layer, the estimation of $\mathbf{c}^T(n)$ is given by $\mathbf{\hat{c}}_{(1)}^T(n)=[\mathbf{\hat{w}}_{(1)}^T(n), \mathbf{\hat{h}}_{(1)}^T(n)] $, according to known signals $y_1(n)=y(n)$ and $\mathbf{u}(n)$. 
The input of the second layer can be obtained by subtracting the SI  
 estimated from the received signal $y(n)$. After SI cancellation, the resultant signal becomes
\begin{equation}
	\begin{aligned}
		y_{(2)}(n)&=\mathbf{c}^T(n)\mathbf{u}(n)-\mathbf{\hat{w}}_{(1)}^T(n)\mathbf{i}(n)+v(n) \\
		&= \mathbf{c}_{(2)}^T(n)\mathbf{u}(n) +v(n)
	\end{aligned}.
\end{equation}		

The same procedure is applied to each subsequent layer of the filter. At the $l$-th layer, it can be represented as 
\begin{equation}
	\begin{aligned}
		y_{(l+1)}(n)&=y_{(l)}(n)\mathbf{-\hat{w}}_{(l)}^T(n)\mathbf{i}(n)+v(n) \\
		&= \mathbf{c}_{(l+1)}^T(n)\mathbf{u}(n) +v(n)
	\end{aligned},
\end{equation}	
where $y_{(l)}(n)$ is the input of the $l$-th layer and $\mathbf{\hat{w}}_{(l)}$ is the estimation of SI channel, and we have $\mathbf{\hat{c}}_{(l)}^T(n)=\mathbf{[\hat{w}}_l^T(n), \mathbf{\hat{h}}_{(l)}^T(n)] $ as the estimation of the joint channel.

 Following a joint estimation process, the SI is estimated and subtracted from the received signal. The resulting residual signal $y_{(l+1)}(n)$ is then utilized as the desired signal for the subsequent layer of filtering in the successive round of joint estimation.	For each level, while eliminating the SI signal, an estimation of the RT channel $\mathbf{\hat{h}}_{(l)}(n)$ is performed. By taking a weighted average of all the estimated values, we can obtain a more accurate estimation of the RT channel as

\begin{equation}
	\begin{aligned}
		\mathbf{\hat{h}}=\frac{1}{L}\sum_{l=1}^L{\hat{\mathbf{h}}_{(l)}}
	\end{aligned},
\end{equation}	
where $L$ represents the number of layers of the filter.

The determination of the number of layers $L$, and the selection of weighted estimated quantities in MTLS is influenced by the ISR. Higher ISR values may necessitate the inclusion of more layers, while excluding the initial estimation $\mathbf{\hat{h}}_{(1)}$ during the weighting process can lead to improved outcomes. Nevertheless, it is important to note that increasing $L$ does not necessarily guarantee enhanced performance once the SI   has been effectively mitigated. Therefore, in order to balance the complexity and performance of the algorithm, $L$ is usually set to 2 or 3.


The detailed procedure of the proposed robust joint channel estimation algorithm is outlined in Algorithm 1, in which, $\mu$ denotes the step size for each layer of MTLS, $N_s $ and $N_w$ represent the length of the training sequence and the sampling window in the M-estimate function, $\hat{c}(n)$ indicates the estimation of $c(n)$, and $({\boldsymbol{A}_e})_{1:M}$ refers to the $1$-th to $M$-th elements of ${\boldsymbol{A}_e}$.  	

	\section{	Simulation Results  }
	
	In this section, we provide the simulation results to verify the effectiveness of the proposed m-MTLS algorithm. 
	We consider a baseband STAR system with a signal bandwidth of 5 kHz. The sampling rate is  set at 10 kHz. Both the local and remote transmitters use the BPSK modulation. 
	The  SI and RT channels have lengths specified as $N=4$ and $M=10$, respectively. And their average energy follows the condition ISR $=E(\|w\|^2/\|h\|^2)$ and SNR $= E(\|s(n)\|^2/\sigma^2_a)$. The received signal $y(n)$ is subject to corruption from two sources: Gaussian white noise $v_a$ and random impulse noise $v_i$. The occurrence probability of the impulse noise is denoted as $P_i$. Additionally, the local reference signal $i(n)$ is influenced by Gaussian noise $v_b$, where the condition $\sigma^2_a=\sigma^2_b$ holds true. For this simulation, the m-MTLS algorithm is utilized with the number of layers $L=3$.
	
	\begin{figure}[!h]
		\label{alg:LSB}
		\renewcommand{\algorithmicrequire}{\textbf{Initialization:}}
		\renewcommand{\algorithmicensure}{\textbf{Output:}}
		\begin{algorithm}[H]
			\caption{m-MTLS algorithm for joint channel estimation}
			\begin{algorithmic}[1]
				\REQUIRE  $\mathbf{\hat{c}}(1)=\mathbf{0}, \hat{\sigma}_e\left( 0 \right) =0$, for $l = 1,...,L$
				\FOR { $n=1,2,...,N_s$ }
				\STATE $y_{(1)}(n)=y(n)={\mathbf{c}^T(n)}\mathbf{u}(n)+v(n)$
				\FOR{$l=1,...,L$}
				\STATE $e_{(l)}(n)=y_{(l)}(n)-{\hat{\mathbf{c}}_{(l)}^T(n)}\mathbf{u}(n)$
				\IF{$n=1$}
				
				\STATE ${\boldsymbol{A}_e}_{(l)}(1)=\underset{N_w}{\underbrace{\left[ e_{(l)}^{2}(1), e_{(l)}^{2}(1) ,..., e_{(l)}^{2}(1) \right] }}$
				\ELSE 
				\STATE ${\boldsymbol{A}_e}_{(l)}(n)=\left[ {{\boldsymbol{A}_e}_{(l)}(n-1)}_{1:N_w-1}, e_{(l)}^{2}(n)\right] $
				
				\ENDIF
				\STATE $\hat{\sigma}^2_e\left( n \right)=\lambda _{\sigma}\hat{\sigma}_{e}^{2}(n-1)+c_2\left( 1-\lambda _{\sigma} \right) \text{med}\left( \boldsymbol{A}_{e(l)}(n) \right) $
				\STATE $\xi =2.576\hat{\sigma}^2_e\left( n \right) $
				\IF{$|e_{(l)}(n)|<\xi$}
				
				\STATE $\mathbf{\hat{c}}_{(l)}(n+1)= \mathbf{\hat{c}}_{(l)}(n) -\mu \hat{\boldsymbol{g}}(\boldsymbol{c}_{(l)}(n))$
				\ELSE
				\STATE $ \mathbf{\hat{c}}_{(l)}(n+1) = \mathbf{\hat{c}}_{(l)}(n)$
				\ENDIF
				\STATE $\mathbf{\hat{w}}_{(l)}(n+1) = {\left(\mathbf{\hat{c}}_{(l)}(n+1)\right)}_{1:N}$
				\STATE $\mathbf{\hat{h}}_{(l)}(n+1) = {\left(\mathbf{\hat{c}}_{(l)}(n+1)\right)}_{N+1:N+M}$
				\STATE  $y_{(l+1)}(n)=y_{(l)}(n)-{\mathbf{\hat{w}}_{(l)}^T(n)}\mathbf{i}(n)$
				\ENDFOR
				\STATE $\mathbf{\hat{h}}(n+1)=\frac{1}{L}\sum_{l=1}^L{\hat{\mathbf{h}}_{(l)}}(n+1)$
				\ENDFOR
			\end{algorithmic}		
		\end{algorithm}
	\end{figure}

\begin{figure}[!ht]
		\centerline{\includegraphics[width=0.78\linewidth]{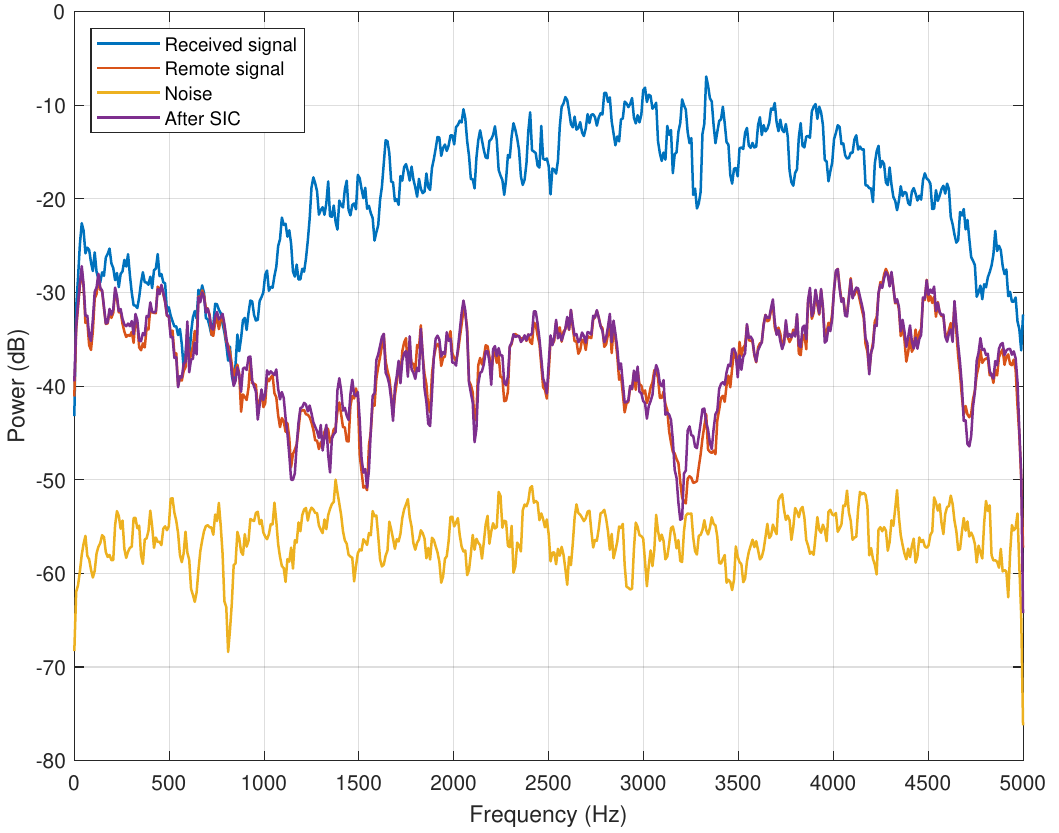}}
		\caption{ Power spectrum of signals at the receiver using m-MTLS with $L = 3$ layers, when ISR $= 20$dB, and SNR $= 20$dB.}
		\label{20db}	
	\end{figure}

\begin{figure}[!ht]
		\centerline{\includegraphics[width=0.78\linewidth]{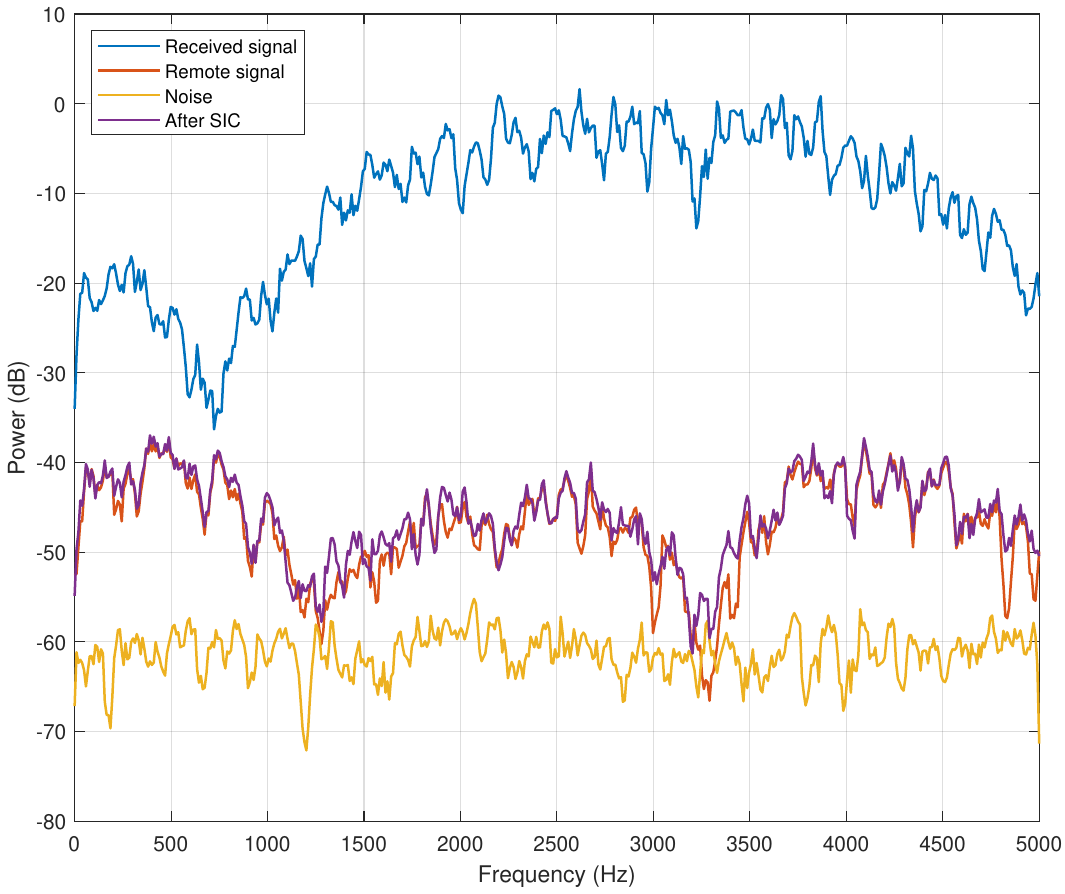}}
		\caption{ Power spectrum of signals at the receiver using m-MTLS with $L = 3$ layers, when ISR $= 40$dB, and SNR $= 20$dB.}
		\label{40db}	
	\end{figure}

	Fig. \ref{20db} and Fig. \ref{40db} compare the power spectrum of different signals at the receiver for ISR $=20$dB and ISR $=40$dB, respectively. It can be observed that in both cases, the received signal power spectrum after SIC is close to the power of the RT signal.  The m-MTLS demonstrates excellent performance by effectively recovering the RT signal after the SIC, even in the high ISR scenarios.
	
	To evaluate the performance of the proposed multi-layered joint estimator, we conduct a comparative analysis with the MMSE estimator and the single-layer MTLS estimator in terms of their efficacy in estimating the RT channel.
	
\begin{figure}[!ht]
		\centerline{\includegraphics[width=0.8\linewidth]{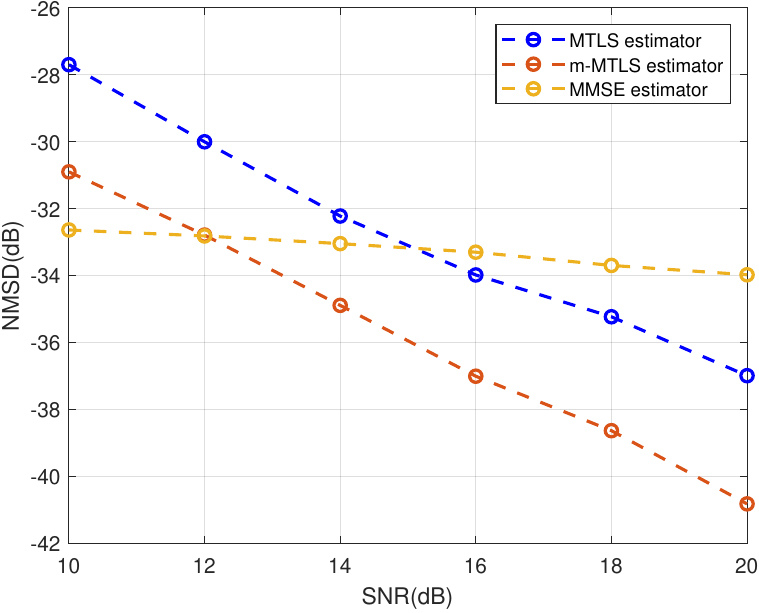}}
		\caption{The performance of different estimators under the impulse noise probability of $P_i=0.01$.}
		\label{001}	
	\end{figure}

	In Fig. \ref{001}, we set  SNR $=20$dB, while the probability of impulse noise is set to $P_i=0.01$. The MMSE estimator exhibits superior performance for low SNR compared to the alternative methods. However, as the SNR increases, it demonstrates limited improvement. Conversely, our proposed m-MTSL estimator showcases notable performance enhancements at high SNR levels, consistently yielding lower NMSD values  than the MTLS and MMSE estimators. 
	
\begin{figure}[!ht]
		\centerline{\includegraphics[width=0.8\linewidth]{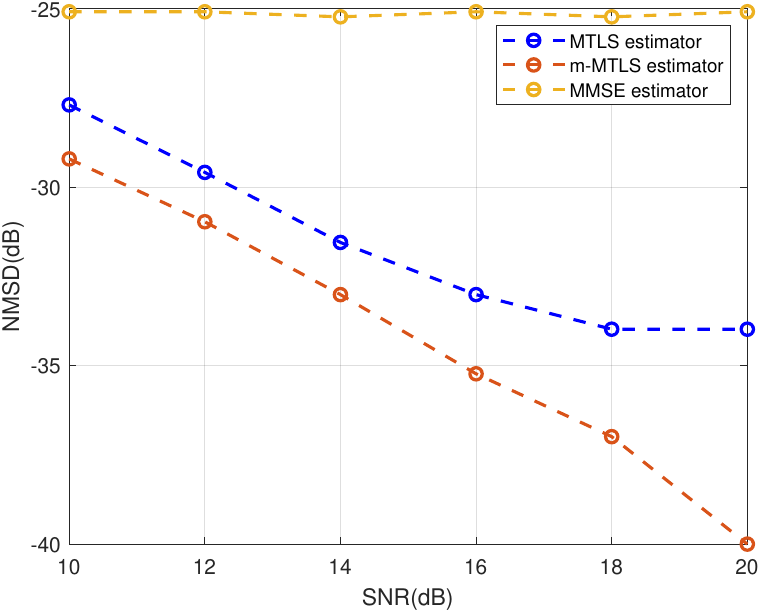}}
		\caption{The performance of different estimators under the impulse noise probability of $P_i=0.05$.}
		\label{005}	
	\end{figure}

	Fig. \ref{005} presents the performance evaluation of the three estimators under an ISR $=20$dB, with an increased probability of impulse noise set to $P_i=0.05$. A comparative analysis of the three curves reveals that m-MTLS consistently achieves superior estimation performance compared to the other two estimators. This improvement  can be attributed to the enhanced robustness facilitated by the MTLS algorithm. Conversely, the MMSE estimator experiences a significant degradation in performance due to the influence of outliers, with this trend becoming more pronounced as the probability of impulse noise increases.
	
	\section{	Conclusion  }
	In this paper, we propose the m-MTLS estimator, which is designed specifically for STAR systems. 
	The m-MTLS estimator effectively addresses the challenges of SIC while accurately estimating the RT channel. 
	Unlike conventional methods that treat the RT signal as noise, our joint estimation approach treats it as part of the input signal and enables more precise estimation of the RT channel and improved cancellation performance, particularly in high SNR scenarios. 
	Although m-MTLS is more complex than MTLS, its structure of multiple layers guarantees accurate estimation even in high ISR conditions.  Additionally, the m-MTLS estimator exhibits impressive performance even in environments with strong impulse noise, thanks to the robustness of the MTLS algorithm.

	\bibliographystyle{ieeetr} 
\bibliography{ref} 
	
\end{document}